\documentclass[12pt,fleqn]{iopart}
\usepackage[english]{babel}
\usepackage{amssymb}
\usepackage{amsfonts}
\usepackage{euscript}

\begin{document}

\title[The 4D Mart{\'{\i}}nez Alonso -- Shabat equation: coverings and recursion operators]
{The four-dimensional Mart{\'{\i}}nez Alonso -- Shabat equation: differential coverings and recursion operators}

\author{Oleg I. Morozov}
\address{Institute of Mathematics and Statistics, University of Troms\o, 
\\
Troms\o
\, 90-37, Norway
\\
Oleg.Morozov{\symbol{64}}uit.no}

\begin{abstract}
We apply Cartan's method of equivalence to find a contact integrable extension
for the structure equations of the symmetry pseudo-group of the four-dimensional Mart{\'{\i}}nez Alonso -- Shabat equation.
From the extension we derive two differential coverings including coverings with one and two non-removable parameters.
Then we apply the same approach to construct a recursion operator for symmetries of the equation under study.
\end{abstract}

\section{Introduction}

We consider the partial differential equation ({\sc pde}) 
\begin{equation}
u_{ty} = u_z\,u_{xy}-u_y\,u_{xz}
\label{rdDym4D}
\end{equation}
introduced by L. Mart{\'{\i}}nez Alonso and A.B. Shabat in   \cite{AlonsoShabat}. This equation has a number of important properties, see, e.g., \cite{BKMV2013,MorozovSergyeyev2014}. In this paper we  apply the technique of \cite{Morozov2009} to find a contact integrable extension
 ({\sc cie}) for the structure equations of the  symmetry pseudo-group of  (\ref{rdDym4D}). This  {\sc cie} produces  two differential coverings, \cite{KrasilshchikVinogradov1989}, for  equation (\ref{rdDym4D}).  In a particular case the first covering contains one non-removable parameter, while in the general case it has two such parameters. To the best of our knowledge there are no examples of coverings with two non-removable parameters in the literature. 

Then we use the  approach of \cite{Morozov2014} to find a B\"acklund autotransformation for the tangent covering,  \cite{KrasilshchikVerbovetsky2011,KrasilshchikVerbovetskyVitolo2012}, of (\ref{rdDym4D}). This transformation provides  a recursion operator for symmetries of (\ref{rdDym4D}).

\section{Coverings of the 4D Mart{\'{\i}}nez Alonso -- Shabat equation}

Using procedures of Cartan's method of equivalence, \cite{Cartan1,Cartan4,Olver1995,FelsOlver,Morozov2002,Morozov2006},  we  compute Maurer--Cartan  forms ({\sc mcf}s) and struc\-tu\-re equations for the symmetry pseudo-group of equation (\ref{rdDym4D}). The struc\-tu\-re equations are given in Appendix. They contain  the following {\sc mcf}s:
\[
\fl
\theta_3 = u_y^{-1} \,\left(
du_y
-(u_z \, u_{xy}-u_y \, u_{xz}) \, dt
-u_{xy} \, dx
-u_{yy} \, dy
-u_{yz} \, dz
\right),
\]
\[
\fl
\theta_4 =
\frac{u_{xy} \, u_{yy}-u_y \, u_{xyy} }{u_y \, R}
\,\left(
du_z
-u_{tz} \, dt
-u_{xz} \, dx
-u_{yz} \, dy
-u_{zz} \, dz
\right)
\]
\[
\fl
\qquad
+
\frac{u_y \, u_{xyz}  -u_{xy} \, u_{yz}}{R}\,\theta_3,
\]
\[
\fl
\xi^1
= \frac{R \, dt}{a \, u_y^2},
\quad
\xi^3
= a \, u_y \,
\left(dy
+ \frac{u_y  \, u_{xyz} - u_{xy} \, u_{yz}}{u_y \,u_{xyy} -u_{xy} \, u_{yy}} \, dz\right),
\quad
\xi^4
=\frac{a \, u_y \, R \, dz}{u_{xy} \,u_{yy}-u_y \,u_{xyy}},
\]
\[
\fl
\xi^2
=
\frac{1}{a\,u_y^3}\,
\left(u_y \,(u_y\, u_{xyz} - u_{xy}\,u_{yz})
-u_z \, (u_y \, u_{xyy}- u_{xy} \, u_{yy}))\, dt
\right.
\]          
\[          %
\fl         %
\qquad  
\left.
+(u_{xy} \, u_{yy}-u_y\,u_{xyy})\, dx\right),
\]
\[
\fl
\eta_4 = \frac{u_y \, du_{xyy} -u_{xy} \, du_{yy}-u_{yy} \, du_{xy}}{u_y  \, u_{xyy} -u_{xy} \, u_{yy}}
-2\,\frac{da}{a}
+ \eta_1  
-\frac{(2\, u_y \, u_{xyy} -3 \, u_{xy} \, u_{yy}) \,d u_y}{u_y\,(u_y \, u_{xyy}-u_{xy} \, u_{yy})},
\]
\begin{equation}
\fl
\eta_1 = \frac{da}{a}+ \frac{1}{u_y} \,\left(u_{xy}\, dx+(u_z \, u_{xy}-u_y \, u_{xz}) \, dt\right),
\label{the_MCFs}
\end{equation}
where  $a \neq 0$ is a parameter and
\[
\fl
R =
(
u_y \, u_{yz} \, u_{xy}^2 \,  (3 \, u_y \, u_{yz} -2 \, u_z \, u_{yy}) \, u_{xyz}
-u_y^2 \, u_{xy} \, (3 \, u_y \, u_{yz}-u_z \, u_{yy}) \, u_{xyz}^2
\]
\[
\fl
\qquad
+u_y^2 \, (u_y\, u_{tzz}  + u_{zz}\, (u_y \, u_{xz}  -u_{z} \, u_{xy}) \, u_{xyy}^2
+u_{xy}^3 \, u_z \, u_{yy} \,(u_{yz}^2 - u_{yy} \,u_{zz}))
\]
\[
\fl
\qquad
+u_y \, u_z \,  u_{xy}^2 \, (u_{yz}^2  -2 \, u_{yy} \, u_{zz}
-2 \, u_y^2 \, u_{xy} \, (u_z \, u_{yz} \, u_{xyz}- u_{xz} \, u_{yy}  \, u_{zz})
)\, u_{xyy}
\]
\[
\fl
\qquad
u_y^3 \, u_z\, u_{xyz}^2\,u_{xyy}
+u_y \, u_{xy}^2 \, (u_{yy}^2 \,u_{tzz}
-u_{xy} \, u_{yz}^3
+u_{xz} \, u_{yy}^2 \, u_{zz})
+u_y^4 \,u_{xyz}^3
)^{1/2}
\cdot
\]
\[
\fl
\qquad
\cdot (
u_z \, (u_y \,u_{xyy}-u_{xy} \, u_{yy})
-3 \, u_y \, (u_y \, u_{xyz}-  u_{xy}\, u_{yz})
)^{-1/2}.
\]
We need no explicit expressions for the other {\sc mcf}s in what follows.

For the structure equations (\ref{SEs_of_rdDym_4D}) we find {\sc cie}s, \cite{Morozov2009}, that have the form
\[
\fl
d \omega_0 = \left(
\sum \limits_{i=0}^4 A_i\,\theta_i
+\sum {}^{*} B_{ij}\,\theta_{ij}
+\sum \limits_{s=1}^{16} C_s\,\eta_s
+\sum \limits_{j=1}^4 D_j\,\xi^j
+\sum \limits_{k=1}^2 E_k\,\omega_k
\right)\wedge \omega_0
\]
\begin{equation}
\fl
\quad\quad\quad
+\sum \limits_{j=1}^4 \left(
\sum \limits_{i=0}^4 F_{ji}\,\theta_i
+\sum \limits_{k=1}^2 G_{jk}\,\omega_k
\right) \wedge \xi^j,
\label{simpliest_CIE}
\end{equation}
where  $\sum {}^{*}$ denotes summation over all $i,j \in \mathbb{N}$ such that
$1\le i \le j \le 4$ and $(i,j)\not = (1,3)$.
The coefficients of (\ref{simpliest_CIE}) are supposed to be either constants or functions of  one additional function $X_0$ with the differential of the form
\begin{equation}
dX_0 =\sum \limits_{i=0}^4 H_i \,\theta_i
+ \sum {}^{*}   I_{ij}\,\theta_{ij}
+ \sum \limits_{s=1}^{16} J_s\,\eta_s
+ \sum \limits_{j=1}^4 K_j\,\xi^j
+ \sum \limits_{q=0}^2 L_q\,\omega_q.
\label{dW_of_simpliest_CIE}
\end{equation}
We require compatibility of 
equations  (\ref{SEs_of_rdDym_4D}) and (\ref{simpliest_CIE}).
This yields an over-determined system of algebraic or ordinary differential equations for the coefficients
of  (\ref{simpliest_CIE}) and (\ref{dW_of_simpliest_CIE}). The analysis of these systems
gives the following result.

\vskip 10 pt 
\noindent
{\bf Theorem 1}. 
{\it 
The structure equations (\ref{SEs_of_rdDym_4D}) have no {\sc cie} (\ref{simpliest_CIE}) with constant coefficients.
Each their {\sc cie} (\ref{simpliest_CIE}), (\ref{dW_of_simpliest_CIE}) with one additional  function $X_0$ is
contact-equivalent to the system
\begin{equation}
\fl
d\omega_0 = (\eta_1 -\omega_1) \wedge \omega_0
+ X_0 \, (\omega_2 +\theta_4) \wedge \xi^1
+ X_0 \,\omega_1 \wedge  \xi^2
      - (\omega_1 +\theta_3) \wedge  \xi^3
+ \omega_2 \wedge  \xi^4,
\label{cie_1}
\end{equation}
\begin{equation}
\fl
dX_0 = X_0   \,(\eta_1 - \eta_4) + Y_1 \,(\omega_0 - X_0\,\xi^2 + \xi^3) + Y_2 \, (X_0\, \xi^1 + \xi^4),
\label{dx_cie_1}
\end{equation}
where $Y_1$, $Y_2$  are arbitrary parameters.
}
\vskip 10 pt 

The inverse third  fundamental Lie theorem in Cartan's form,  \cite[\S\S 16--24]{Cartan1}, \cite{Cartan4},
\cite[\S\S 16, 19, 20, 25, 26]{Vasilieva1972}, \cite[\S\S 14.1--14.3]{Stormark2000}, ensures the existence of the
forms $\omega_0$, $\omega_1$, $\omega_2$ and the function $X_0$ satisfying system (\ref{cie_1}), (\ref{dx_cie_1}).
Since the {\sc mcf}s  (\ref{the_MCFs})  are known explicitly, it is not hard to find the form $\omega_0$. The
integration of (\ref{cie_1}), (\ref{dx_cie_1}) depends on whether the conditions $Y_1 \equiv 0$ or  $Y_1 \not \equiv 0$ hold.
In the first case we have 
\begin{equation}
\fl
\omega_0 = a \,H^{-1} \, q_x^{-1} \,\left(
dq -  (u_z \, q_x - H^{-1}\,q_z) \,dt - q_x \, dx - H\, u_y \, q_x \, dy - q_z \,dz
\right),
\label{WE_form_H}
\end{equation}
where $H = H(t,z)$ is a solution to equation
\begin{equation}
H_z + H \,H_t = 0,
\label{H_eq}
\end{equation}
while in the case of $Y_1 \not \equiv 0$ 
\[
\fl
\omega_0 = a \, q^{-1} \, q_x^{-1} \,\left(
dq -  (u_z \, q_x - q^{-1}\,q_z) \,dt - q_x \, dx - q \, u_y \, q_x \, dy - q_z \,dz
\right).
\]

\vskip 10 pt
\noindent
{\bf Corollary 1}.
{\it
Equation (\ref{rdDym4D}) has two coverings defined by sys\-tems
\begin{equation}
\left\{
\begin{array}{lcl}
q_t& =& u_z \, q_x - H^{-1}\,q_z,
\\
q_y &=& H \, u_y \, q_x,
\end{array}
\right.
\label{H_covering}
\end{equation}
where $H$ is a solution to equation (\ref{H_eq}),  and
\[
\left\{
\begin{array}{lcl}
q_t& =& u_z \, q_x - q^{-1}\,q_z,
\\
q_y &=&  u_y \,q \, q_x.
\end{array}
\right.
\]
}

\vskip 10 pt

Equation (\ref{H_eq})  has constant solutions $H(t,z) \equiv \lambda = \mathrm{const}$. The parameter $\lambda$ in the corresponding system
\begin{equation}
\left\{
\begin{array}{lcl}
q_t& =& u_z \, q_x - \lambda^{-1}\,q_z,
\\
q_y &=& \lambda \, u_y \, q_x
\end{array}
\right.
\label{first_covering}
\end{equation}
is non-removable. In accordance with \cite[\S\S 3.2, 3.6]{KrasilshchikVinogradov1989}, \cite{Krasilshchik2000,IgoninKrasilshchik2002,Marvan2002}
to prove this claim it is enough to show that it is impossible to lift the symmetry $V =  t\, \frac{\partial}{\partial t} - u\, \frac{\partial}{\partial u}$
of equation (\ref{rdDym4D})  to a symmetry of system  (\ref{first_covering}), while the action of
$\mathrm{exp}\left(\mathrm{ln} (\lambda) \,V\right)$ on the form (\ref{WE_form_H}) with $H = 1$ gives (\ref{WE_form_H}) with  $H=\lambda$.

When $H \not \equiv \mathrm{const}$, the symmetries $\frac{\partial}{\partial t}$ and $\frac{\partial}{\partial z}$ can not be lifted to symmetries of system (\ref{H_covering}). Then for each solution $H(t,z)$ of (\ref{H_eq})  we have a two-parameter family of coverings  
\[
\left\{
\begin{array}{lcl}
q_t& =& u_z \, q_x - (H(t+\lambda,z+\mu))^{-1}\,q_z,
\\
q_y &=& H(t+\lambda,z+\mu) \, u_y \, q_x,
\end{array}
\right.
\]
with non-removable parameters $\lambda$ and $\mu$. For example, solution $H = t\,z^{-1}$ produces the covering
\[
\left\{
\begin{array}{lcl}
q_t& =& u_z \, q_x - (z+\mu)\,(t+\lambda)^{-1}\,q_z,
\\
q_y &=& (t+\lambda)\,(z+\mu)^{-1}\, u_y \, q_x.
\end{array}
\right.
\]

\section{A recursion operator for symmetries of the 4D Mart{\'{\i}}nez Alonso -- Shabat equation}

The generators of infinitesimal symmetries, \cite{Vinogradov1984,KrasilshchikLychaginVinogradov1986,Olver1993}, of equation (\ref{rdDym4D}) are solutions to the restriction of its {\it linearization}
\begin{equation}
\fl
D_t \circ D_y(\varphi) = u_z\,D_x \circ D_y(\varphi)-u_y\,D_x \circ D_z(\varphi)
+ u_{xy} \,D_z(\varphi) -u_{xz} \,D_y(\varphi)
\label{linearized_rdDym4D}
\end{equation}
on (\ref{rdDym4D}). Consider the associated to (\ref{linearized_rdDym4D}) equation
\begin{equation}
v_{ty} = u_z\,v_{xy}-u_y\,v_{xz}+ u_{xy} \,v_z -u_{xz} \,v_y.
\label{tangent_rdDym4D}
\end{equation}
The structure equations of the symmetry pseudo-group of system (\ref{rdDym4D}), (\ref{tangent_rdDym4D}) include equations  (\ref{SEs_of_rdDym_4D}) and the involutive completion of equations
\[
\fl
d\zeta_0
=
(\eta_{1} -    \zeta_3+X_3 \, \xi^{1}+X_2 \, \xi^{2}+X_1 \, \xi^{3}+X_4 \, \xi^{4} ) \wedge \zeta_0
-\zeta_1 \wedge \xi^{1}
-\zeta_2 \wedge \xi^{2}
-\zeta_3 \wedge \xi^{3}
\]
\[
\fl
\qquad
-\zeta_4 \wedge \xi^{4}
+\theta_{0} \wedge (\zeta_3  -X_1 \, \xi^{3}-X_4 \, \xi^{4}),
\]
\[
\fl
d\zeta_1
=
-\theta_{24} \wedge \zeta_0
+(\eta_{1} -\eta_{2} - \zeta_3) \wedge \zeta_1
-\eta_{3} \wedge \zeta_2
+(\eta_{34} + X_1 \, (\zeta_1+\theta_{1})) \wedge \theta_{0}
-\theta_{1} \wedge \zeta_3
\]
\[
\fl
\qquad
+\eta_{32} \wedge \xi^{1}
+\eta_{33} \wedge \xi^{2}
+\eta_{34} \wedge \xi^{3}
+\eta_{35} \wedge \xi^{4},
\]
\[
\fl
d\zeta_2
=
\zeta_1 \wedge (X_2 \, \xi^{1}-X_1 \, \xi^{4})
+(\eta_{1}-\eta_{4} + X_3 \, \xi^{1}-X_1 \, \theta_{0}+X_4 \, \xi^{4})\wedge \zeta_2
+\zeta_0 \wedge (\theta_{0}+\theta_{23})
\]
\[
\fl
\qquad
+(\zeta_2 +\theta_{2}) \wedge \zeta_3
+(\eta_{37} + X_1 \, \theta_{2})\wedge \theta_{0}
+\eta_{33} \wedge \xi^{1}
+\eta_{36} \wedge \xi^{2}
+\eta_{37} \wedge \xi^{3}
\]
\[
\fl
\qquad
+(\eta_{24}-\eta_{34}-X_1 \, \theta_{1}-X_4 \, \theta_{2}-X_5 \, (\theta_{23}+\theta_{0}))\wedge \xi^{4},
\]
\[
\fl
d\zeta_3
=
(\eta_{34} + X_1 \, (\zeta_1 + \theta_{1})- X_3 \, \zeta_3  )\wedge\xi^{1}
+(\eta_{37}+\zeta_0 + X_1 \, (\zeta_2 + \theta_{2}) - X_3 \, \zeta_3)\wedge \xi^{2}
\]
\[
\fl
\qquad
+\eta_{38} \wedge \xi^{3}
+\eta_{39} \wedge \xi^{4},
\]
\[
\fl
d\zeta_4
=
(\eta_{3} - X_4 \, \xi^{3}) \wedge \zeta_3
-
(\zeta_3 + \eta_{2} - \eta_{4}-X_3 \, \xi^{1}-X_2 \, \xi^{2}-X_1 \, \xi^{3}) \wedge \zeta_4
\]
\[
\fl
\qquad
+
(\eta_{35}-\zeta_0 + X_4 \,(\zeta_1+\theta_{1}))\wedge \xi^{1}
-
(\eta_{34} + X_1 \,(\zeta_1+\theta_{1})+X_5 \, \theta_{23}-\theta_{24})\wedge \xi^{2}
\]
\begin{equation}
\fl
\qquad
+
(\eta_{39}+ X_4 \,\theta_{3} -X_1 \,\theta_{4})\wedge \xi^{3}
+
\eta_{40} \wedge \xi^{4}.
\label{SEs_of_tangent_rdDym_4D}
\end{equation}
The involutive completion of (\ref{SEs_of_tangent_rdDym_4D}) contains also equations for the differentials
$d \eta_{32}$, ... , $d \eta_{40}$, but they are too big to be written in full here.
The invariants $X_1$, ... , $X_7$ in (\ref{SEs_of_tangent_rdDym_4D}) have the following differentials:
\[
\fl
d X_1 =\eta_{38}-X_1 \, (\eta_{1}+\theta_{3} - X_3 \, \xi^{1} - X_2 \, \xi^{2}),
\]
\[
\fl
dX_2
=
\eta_{37}
+\zeta_0
+X_1 \, (\zeta_2+\theta_{2}+X_2 \, \xi^{3})
-X_2 \, (\zeta_3+\eta_{4})
+(X_5+X_2 \, X_4) \, \xi^{4},
\]
\[
\fl
d X_3
=
\eta_{34}
+X_1 \, (\zeta_1+\theta_{1})
-X_2 \, \eta_{3}
-X_3 \, (\zeta_3+\eta_{2}- X_4 \, \xi^{4}- X_1 \, \xi^{3}),
\]
\[
\fl
d X_4
=
\eta_{39}
+X_1 \, (\eta_{3}-\theta_{4})
-X_4 \, (\eta_{1}+\eta_{2}-\eta_{4}-X_3 \, \xi^{1})
+(X_5+X_2 \, X_4) \, \xi^{2},
\]
\[
\fl
d X_5
=
\zeta_4
+\eta_{3}
-X_5 \, (\zeta_3+\eta_{2}-\eta_{4})
+X_7 \, \xi^{1}
+(X_2 \, X_5+X_3) \, \xi^{2}
+(X_1 \, X_5-X_4) \, \xi^{3}
+X_6 \, \xi^{4},
\]
\[
\fl
d X_6
=
-\eta_{40}
+X_5 \, \eta_{39}
-X_6 \, (\zeta_3+\eta_{1}+2 \, \eta_{2}-2 \,\eta_{4}-X_1 \, \xi^{3})
+(X_1 \, X_5-X_4) \, (\eta_{3}-\theta_{4})
\]
\[
\fl
\qquad
+(X_5^2+X_2 \, (X_6+ X_4 \,X_5)+X_3 \, X_4+1) \, \xi^{2},
\]
\[
\fl
d X_7 =
\zeta_0
-\eta_{35}
+\theta_{14}
+X_3 \, \zeta_4
+X_5 \, (\theta_{24}+\eta_{34}-\xi^{4})
+X_7 \, (\eta_{4}-2 \, \eta_{2} +X_2 \, \xi^{2})
\]
\[
\fl
\qquad
+(X_1 \, X_5-X_4) \, (\zeta_1+\theta_{1})
-(X_7+X_3 \, X_5) \, \zeta_3
-(X_3+X_2 \, X_5) \, \eta_{3}
\]
\[
\fl
\qquad
+(X_1 \, X_7+X_3 \,(X_1 \, X_5-X_4)+1) \, \xi^{3}.
\]
The {\sc mcf}s $\zeta_0$, ... , $\zeta_4$ in (\ref{SEs_of_tangent_rdDym_4D}) are
\[
\fl
\zeta_0  =  a \, u_y\, v_y^{-1} (d v - v_t \,dt - v_x \,dx - v_y \,dy - v_z \,dz) - \theta_0,
\]
\[
\fl
\zeta_1  = a^2 \, u_y^3\,v_y^{-1}\,R^{-1}\,  (d v_t - v_{tt} \,dt - v_{tx} \,dx - v_{ty} \,dy - v_{tz} \,dz)
+a \, u_y\,(u_y u_{xz}-u_z u_{xy})\,R^{-1}\,\zeta_0
\]
\[
\fl
\qquad
- (u_y^2 u_{xyz}-u_y u_z u_{xyy} + u_{xy}\,(u_z u_{yy}-u_y u_{yz}))\, u_y^{-1} R^{-1} \,(\zeta_2 + \theta_2)
- \theta_1,
\]
\[
\fl
\zeta_2 = a \, u_y^2 v_y \,(u_y u_{xyy}-u_{xy} u_{yy})^{-1}\,
(a\,u_y^2 \,(d v_x - v_{tx} \,dt - v_{xx} \,dx - v_{xy} \,dy - v_{xz} \,dz)
 + u_y v_{xy}\, \zeta_0
\]
\[
\fl
\qquad
  + u_{xy} v_y \,\theta_0 - \theta_2,
\]
\[
\fl
\zeta_3 = v_y^{-1} \, (d v_y - v_{ty} \,dt - v_{xy} \,dx - v_{yy} \,dy - v_{yz} \,dz) - \theta_3,
\]
\[
\fl
\zeta_4 = (u_{xy} u_{yy} - u_y u_{yyy})\,v_y^{-1}R^{-1}\, (d v_z - v_{tz} \,dt - v_{xz} \,dx - v_{yz} \,dy - v_{zz} \,dz)
- \theta_4,
\]
where $v_{ty}$ is replaced by the right-hand side of (\ref{tangent_rdDym4D}).

\vskip 10 pt
The analysis of {\sc cie}s of system (\ref{SEs_of_rdDym_4D}), (\ref{SEs_of_tangent_rdDym_4D}) gives the following result.

\vskip 10 pt
\noindent
{\bf Theorem 2}.
{\it 
The structure equations 
(\ref{SEs_of_rdDym_4D}),
(\ref{SEs_of_tangent_rdDym_4D})
have a {\sc cie}
\[
\fl
d\zeta_5
=
X_8 \,  (\xi^2 - X_5 \,\xi^1) \wedge  \zeta_5
+ (\theta_{24} - X_8 \, \zeta_4) \wedge \xi^1
+ (X_8 \, \zeta_3 - \theta_{23}) \wedge \xi^2
+ \zeta_{6}  \wedge \xi^3
\]
\begin{equation}
\fl
\qquad
+ \zeta_{7} \wedge \xi^4
\label{cie_of_tangent_rdDym4D}
\end{equation}
with one additional invariant $X_8$, whose differential satisfies equation
\begin{equation}
\fl
d X_8
=
X_8 \, (\zeta_3 - \zeta_5 - \eta_4 - (X_3 + X_5 \, X_8) \, \xi^1 - (X_2 - X_8) \, \xi^2)
+ Y_3 \, \xi^3
+ Y_4 \, \xi^4,
\label{dX_in_cie_of_tangent_rdDym4D}
\end{equation}
with arbitrary parameters $Y_3$, $Y_4$.
}

Applying the inverse  third fundamental Lie theorem to system (\ref{cie_of_tangent_rdDym4D}),
(\ref{dX_in_cie_of_tangent_rdDym4D}) we find
\[
\fl
\zeta_5
=
\frac{1}{w}\,
\left(d w
-\frac{u_y \, v_z - u_z \, v_y  +  (u_z \, u_{xy}-u_y\,u_{xz}) \, w   }{u_y} \, dt
-\frac{u_{xy} \, w -v_y}{u_y} \, dx
-w_y \, dy-w_z \, dz
\right).
\]
This form defines the covering
\begin{equation}
\left\{
\begin{array}{lcl}
w_t &=& \displaystyle{
u_y^{-1} \, \left(
u_y \, v_z - u_z\, v_y + (u_z\,u_{xy}-u_y\,u_{xz})\, w \right),
}
\\
w_x &=& \displaystyle{
u_y^{-1} \, \left( u_{xy} \,w - v_y\right)}
\end{array}
\right.
\label{covering_for_tangent_rdDym4D}
\end{equation}
for system (\ref{rdDym4D}), (\ref{tangent_rdDym4D}).
Exclusion of  $v$ from  (\ref{covering_for_tangent_rdDym4D}) shows that $w$ is also a solution to (\ref{tangent_rdDym4D}).  Therefore (\ref{covering_for_tangent_rdDym4D}) is a B\"acklund auto-transformation
of system (\ref{rdDym4D}), (\ref{tangent_rdDym4D}).  In other words system
\[
\left\{
\begin{array}{lcl}
D_t(\psi) &=& \displaystyle{
u_y^{-1} \, \left(
u_y \, D_z(\varphi) - u_z\, D_y(\varphi) + (u_z\,u_{xy}-u_y\,u_{xz})\, \psi \right),
}
\\
D_x(\psi) &=& \displaystyle{
u_y^{-1} \, \left( u_{xy} \,\psi - D_y(\varphi)\right)}
\end{array}
\right.
\]
defines a recursion operator $\EuScript{R}$ for symmetries of equation (\ref{rdDym4D}): if $\varphi$ is a solution to
(\ref{linearized_rdDym4D}), then $\psi = \EuScript{R}(\varphi)$ is a solution to (\ref{linearized_rdDym4D}) as well.
The inverse operator $\varphi = \EuScript{R}^{-1}(\psi)$ is defined by system
\[
\left\{
\begin{array}{lcl}
D_y(\varphi) &=& -u_y \, D_x(\psi) + u_{xy} \, \psi,
\\
D_z(\varphi) &=& D_t(\psi) - u_z \, D_x(\psi) + u_{xz} \, \psi.
\end{array}
\right.
\]

\section*{Acknowledgments}

I'm very grateful to I.S. Krasil${}^{\prime}$shchik, A.G. Sergyeyev and M.V. Pavlov for enlightening and stimulating discussions. I thanks A.G. Sergyeyev for the warm hospitality in the Silesian University in Opava, where this work was initiated and partially supported by the Grant Agency of the Czech Republic (GA ${\check{\mathrm{C}}}$R) under grant P201/12/G028.

\section*{Appendix}
The structure equations of the symmetry pseudo-group of eqaution (\ref{rdDym4D}) read
{\small
\[
\fl
d\theta_{0} =
\eta_{1} \wedge \theta_{0}
+ \xi^1 \wedge \theta_{1}
+ \xi^2 \wedge \theta_{2}
+ \xi^3 \wedge \theta_{3}
+ \xi^4 \wedge \theta_{4},
\]
\[
\fl
d\theta_{1} =
(\eta_{1}-\eta_{2}) \wedge \theta_{1}
-\eta_{3} \wedge \theta_{2}
+\theta_{0} \wedge (\xi^4-\theta_{24})
+\xi^1 \wedge \theta_{11}
+\xi^2 \wedge \theta_{12}
-\xi^3 \wedge \theta_{24}
+\xi^4 \wedge \theta_{14},
\]
\[
\fl
d\theta_{2} =
(\eta_{1} -\eta_{4}) \wedge \theta_{2}
+\theta_{0} \wedge (\xi^3-\theta_{23})
+\xi^1 \wedge \theta_{12}
+\xi^2 \wedge \theta_{22}
+\xi^3 \wedge \theta_{23}
+\xi^4 \wedge \theta_{24},
\]
\[
\fl
d\theta_{3} =
-\xi^1 \wedge \theta_{24}
+\xi^2 \wedge \theta_{23}
+\xi^3 \wedge \theta_{33}
+\xi^4 \wedge \theta_{34},
\]
\[
\fl
d\theta_{4} =
(\eta_{4} - \eta_{2})\wedge \theta_{4}
+\eta_{3} \wedge \theta_{3}
+\theta_{3} \wedge \theta_{4}
+\xi^1 \wedge \theta_{14}
+\xi^2 \wedge \theta_{24}
+\xi^3 \wedge \theta_{34}
+\xi^4 \wedge \theta_{44},
\]
\[
\fl
d\xi^1 = \eta_{2} \wedge \xi^1,
\]
\[
\fl
d\xi^2 =
\eta_{3} \wedge \xi^1
+\eta_{4} \wedge \xi^2,
\]
\[
\fl
d\xi^3 =
(\eta_{1} +\theta_{3})\wedge \xi^3
-(\eta_{3} - \theta_{4})\wedge \xi^4,
\]
\[
\fl
d\xi^4 =
(\eta_{1} +\eta_{2} -\eta_{4}) \wedge \xi^4,
\]
\[
\fl
d\theta_{11} =
(\eta_{1}-2 \,\eta_{2}) \wedge \theta_{11}
-2 \,\eta_{3} \wedge  \theta_{12}
+\eta_{9} \wedge \xi^2
-\eta_{10} \wedge (\theta_{0}+\xi^3)
+\eta_{11} \wedge \xi^1
+\theta_{2} \wedge (\xi^3+\theta_{14})
\]          
\[          %
\fl         %
\qquad  
+\eta_{12} \wedge \xi^4
-2\,\theta_{1} \wedge  \theta_{24},
\]
\[
\fl
d\theta_{12} =
(\eta_{1} -\eta_{2} -\eta_{4}) \wedge \theta_{12}
-\eta_{3} \wedge \theta_{22}
+\eta_{5} \wedge (\theta_{0}+\xi^3)
+\eta_{7} \wedge \xi^2
+\eta_{9} \wedge \xi^1
+\eta_{10} \wedge \xi^4
\]          
\[          %
\fl         %
\qquad  
+\theta_{1} \wedge (\theta_{23}-\xi^3),
\]
\[
\fl
d\theta_{14} =
(\eta_{4} -2\, \eta_{2})\wedge (\xi^3+\theta_{14})
-(\eta_{1} +\theta_{3})\wedge \xi^3
+\eta_{3} \wedge (3\, \xi^4-2\, \theta_{24})
+\eta_{10} \wedge \xi^2
-\theta_{4} \wedge \xi^4
\]          
\[          %
\fl         %
\qquad  
+(\eta_{12} -2\,\theta_{1})\wedge \xi^1,
\]
\[
\fl
d\theta_{22} =
(\eta_{1} -2\,\eta_{4}) \wedge  \theta_{22}
-\eta_{5} \wedge \xi^4
+\eta_{6} \wedge (\theta_{0}+\xi^3)
+\eta_{7} \wedge \xi^1
+\eta_{8} \wedge \xi^2
+\theta_{2} \wedge (\theta_{23}-\xi^3),
\]
\[
\fl
d\theta_{23} =
(\eta_{1} +\theta_{3})\wedge \xi^3
-(\eta_{3} - \theta_{4})\wedge \xi^4
+\eta_{4} \wedge (\xi^3-\theta_{23})
+\eta_{5} \wedge \xi^1
+\eta_{6} \wedge \xi^2,
\]
\[
\fl
d\theta_{24} =
(\eta_{1} -\eta_{4})\wedge \xi^4
+\eta_{2} \wedge (2\, \xi^4-\theta_{24})
+\eta_{3} \wedge (\theta_{23}-\xi^3)
-\eta_{5} \wedge \xi^2
+\eta_{10} \wedge \xi^1,
\]
\[
\fl
d\theta_{33} =
(\eta_{1} +\theta_{3})\wedge (\xi^2-\theta_{33})
+\eta_{3} \wedge \xi^1
+\eta_{4} \wedge \xi^2
+\eta_{13} \wedge \xi^3
+\eta_{14} \wedge \xi^4,
\]
\[
\fl
d\theta_{34} =
(\eta_{4}-\eta_{1}) \wedge (\xi^1+\theta_{34})
-\eta_{2} \wedge (2\, \xi^1+\theta_{34})
-(\eta_{3} - \theta_{4})\wedge (\xi^2-\theta_{33}),
\]
\[
\fl
d\theta_{44} =
(\eta_{1} -\theta_{3}) \wedge (\xi^2-\theta_{44})
+(\eta_{2} -\eta_{4}) \wedge 2\, (\xi^2-2\, \theta_{44})
+\eta_{3} \wedge (3\, \xi^1+2\, \theta_{34})
+\eta_{15} \wedge \xi^3
\]          
\[          %
\fl         %
\qquad  
+\eta_{16} \wedge \xi^4
-2\,\theta_{4} \wedge (\xi^1+\theta_{34}),
\]
\[
\fl
d\eta_{1} =
\xi^1 \wedge (\theta_{24} -\xi^4)
-\xi^2 \wedge (\theta_{23}-\xi^3),
\]
\[
\fl
d\eta_{2} = 0 ,
\]
\[
\fl
d\eta_{3} =
(\eta_{4}-\eta_{2}) \wedge \eta_{3}
+\xi^1 \wedge (\xi^3+\theta_{14})
-\xi^2 \wedge (\xi^4-\theta_{24}),
\]
\[
\fl
d\eta_{4} =
\xi^1 \wedge (\theta_{24} -\xi^4)
-\xi^2 \wedge (\theta_{23}-\xi^3),
\]
\[
\fl
d\eta_{5} =
\eta_{5} \wedge (\eta_{2}+\eta_{4})
-\eta_{3} \wedge \eta_{6}
+\eta_{17} \wedge \xi^1
+\eta_{18} \wedge \xi^2
+(\theta_{23} - \xi^3) \wedge(\xi^4 -\theta_{24}),
\]
\[
\fl
d\eta_{6} =
2\,\eta_{6} \wedge\eta_{4}
+\eta_{18} \wedge \xi^1
+\eta_{19} \wedge \xi^2,
\]
\[
\fl
d\eta_{7} =
(\eta_{1} -\eta_{2} -2\, \eta_{4})\wedge \eta_{7}
-\eta_{3} \wedge \eta_{8}
-\eta_{5} \wedge \theta_{2}
-\eta_{6} \wedge \theta_{1}
-(\eta_{17} - \theta_{22}) \wedge \xi^4
+\eta_{18} \wedge (\theta_{0}+\xi^3)
\]          
\[          %
\fl         %
\qquad  
+\eta_{20} \wedge \xi^1
+\eta_{21} \wedge \xi^2
+\theta_{12} \wedge (\xi^3-\theta_{23})
-\theta_{22} \wedge \theta_{24},
\]
\[
\fl
d\eta_{8} =
(\eta_{1} -3\, \eta_{4}) \wedge \eta_{8}
-2\, \eta_{6} \wedge  \theta_{2}
-\eta_{18} \wedge \xi^4
+\eta_{19} \wedge (\theta_{0}+\xi^3)
+\eta_{21} \wedge \xi^1
+\eta_{22} \wedge \xi^2,
\]
\[
\fl
d\eta_{9} =
(\eta_{1} -2\, \eta_{2}-\eta_{4})\wedge \eta_{9}
-2\, \eta_{3} \wedge \eta_{7}
-2\, \eta_{5} \wedge \theta_{1}
+\eta_{17} \wedge (\theta_{0}+\xi^3)
+\eta_{23} \wedge \xi^1
+\eta_{20} \wedge \xi^2
\]          
\[          %
\fl         %
\qquad  
+\eta_{24} \wedge \xi^4
+(\theta_{11}-\theta_{22})  \wedge \xi^3
-\theta_{11} \wedge \theta_{23}
+\theta_{14} \wedge \theta_{22},
\]
\[
\fl
d\eta_{10} =
2\,\eta_{10} \wedge \eta_{2}
+2\,\eta_{3} \wedge \eta_{5}
-\eta_{17} \wedge \xi^2
+\eta_{24} \wedge \xi^1
+(\theta_{14}+\theta_{23}) \wedge \xi^3
-\theta_{14} \wedge\theta_{23},
\]
\[
\fl
d\eta_{11} =
(\eta_{1} -3\,\eta_{2})\wedge \eta_{11}
-3\, \eta_{3} \wedge \eta_{9}
+3\,\eta_{10} \wedge  \theta_{1}
-(\eta_{12} - 2\,\theta_{1})\wedge \theta_{2}
+\eta_{23} \wedge \xi^2
+\eta_{25} \wedge \xi^1
\]          
\[          %
\fl         %
\qquad  
-\eta_{24} \wedge (\theta_{0}+\xi^3)
+\eta_{26} \wedge \xi^4
-3\,\theta_{12} \wedge  (\theta_{14} - \xi^3)
+3\,\theta_{11} \wedge  \theta_{24},
\]
\[
\fl
d\eta_{12} =
2\, \eta_{1} \wedge  \theta_{1}
+\eta_{2} \wedge (4\, \theta_{1}-3\, \eta_{12})
-\eta_{3} \wedge (2\, \theta_{2}+3\, \eta_{10})
+\eta_{4} \wedge (\eta_{12}-2\, \theta_{1})
+\eta_{24} \wedge \xi^2
\]          
\[          %
\fl         %
\qquad  
+\eta_{26} \wedge \xi^1
+2\,\theta_{0} \wedge  (\xi^4-\theta_{24})
- \xi^1 \wedge \theta_{11}
+2\,\xi^2 \wedge  \theta_{12}
+\xi^3 \wedge (\theta_{24}-3\, \xi^4)
\]          
\[          %
\fl         %
\qquad  
+\theta_{14} \wedge (3\, \theta_{24}-5\,\xi^4),
\]
\[
\fl
d\eta_{13} =
2\,\eta_{13} \wedge  (\eta_{1}+\theta_{3})
+\eta_{27} \wedge \xi^3
+\eta_{28} \wedge \xi^4,
\]
\[
\fl
d\eta_{14} =
(\eta_{4} -2\,\eta_{1} - \eta_{2}-\theta_{3}) \wedge \eta_{14}
+(\eta_{3} - \theta_{4})\wedge \eta_{13}
+\eta_{28} \wedge \xi^3
+\eta_{29} \wedge \xi^4
\]          
\[          %
\fl         %
\qquad      
+(\theta_{33}-\xi^2) \wedge (\theta_{34}- \xi^1),
\]
\[
\fl
d\eta_{15} =
2\,(\eta_{4}-\eta_{1}-\eta_{2}) \wedge \eta_{15}
+2\,(\eta_{3}-\theta_{4}) \wedge 2\, \eta_{14}
+\eta_{29} \wedge \xi^3
+\eta_{30} \wedge \xi^4
+\xi^2 \wedge (\theta_{44}-\theta_{33})
\]          
\[          %
\fl         %
\qquad      
-\theta_{33} \wedge \theta_{44},
\]
\[
\fl
d\eta_{16} =
(3\,\eta_{4} -2\, \eta_{1} -3\, \eta_{2} +\theta_{3}) \wedge \eta_{16}
+3\, (\eta_{3}- \theta_{4})\wedge  \eta_{15}
+\eta_{30} \wedge \xi^3
+\eta_{31} \wedge \xi^4
\]          
\begin{equation}          %
\fl         %
\qquad      
+3\, (\theta_{34}+\xi^1) \wedge (\xi^2-\theta_{44}).
\label{SEs_of_rdDym_4D}
\end{equation}

}

\end{document}